\documentclass{kapproc} 

\usepackage{procps} 

\usepackage[dvips]{graphicx}

\upperandlowercase

\kluwerbib 

\begin{document}

\articletitle[X-Ray Sources Overdensity Around the 3C 295 Galaxy Cluster]
{X-Ray Sources Overdensity Around 3C 295}

\author{V. D'Elia, F. Fiore, M. Elvis, M. Cappi, S. Mathur, P. Mazzotta
  \& E. Falco}

\anxx{Beerends\, John G.}

\begin{abstract}
We present a statistical analysis of the Chandra observation of 
the source field around the 3C 295 galaxy cluster ($z=0.46$).  
Three different methods of analysis, namely a chip by chip 
logN-logS, a two dimentional Kolmogorov-Smirnov (KS) test, and the angular
correlation function (ACF) show a strong overdensity of
sources in the North-East of the field, that
may indicate a filament of the large scale structure of the Universe 
toward 3C 295. 

\end{abstract}

\section{Observation and Data Reduction}

Chandra observed the $16' \times 16'$ field around the 
3C 295 cluster with ACIS-I on May 18, 2001, for $92$ ks. 
All the analysis has been performed separately in the $0.5-2$ keV
($89$ sources identified),
in the $2-7$ keV ($71$ sources) and in the $0.5-7$ keV ($121$
sources, fig. 1a) bands. 
The counts in the three bands were converted in 
$0.5 -2$ keV, $2 -10$ keV and $0.5 -10$ keV fluxes, respectively.

\section{Main Results}

The following results have been achieved: 

3C 295 LogN-LogS  is consistent with the 
CDFS LogN-LogS by Rosati et al. 2002 in both the soft and hard band.
The 3C 295 LogN-LogS in the soft, hard and broad bands computed
separately for each ACIS-I chip show an overdensity of sources in the
North-East (NE) chip (fig. 1b) which reflects the clustering of sources
clearly visible in fig. 1.
The discrepancy between the normalization of the LogN-LogS for the NE
and SW chip is $3.2\sigma$, $3.3\sigma$ and $4.0\sigma$ in the soft, 
hard and broad band, respectively (fig. 2a). 

The two dimensional KS test shows the probability that the
3C 295 sources are uniformly distributed is only $\sim 3\%$ in the
soft and hard bands, and drops below 1\% in the broad band.  

The ACF shows a strong signal
on scales of a few arcmins ($\sim$ half a chip length), and
also on lower scales in the $0.5 - 7$ keV band (fig. 2b).

More details on the present work can be found in D'Elia et al., A\&A, submitted.

\begin{figure}[ht]
\includegraphics[scale=0.34]{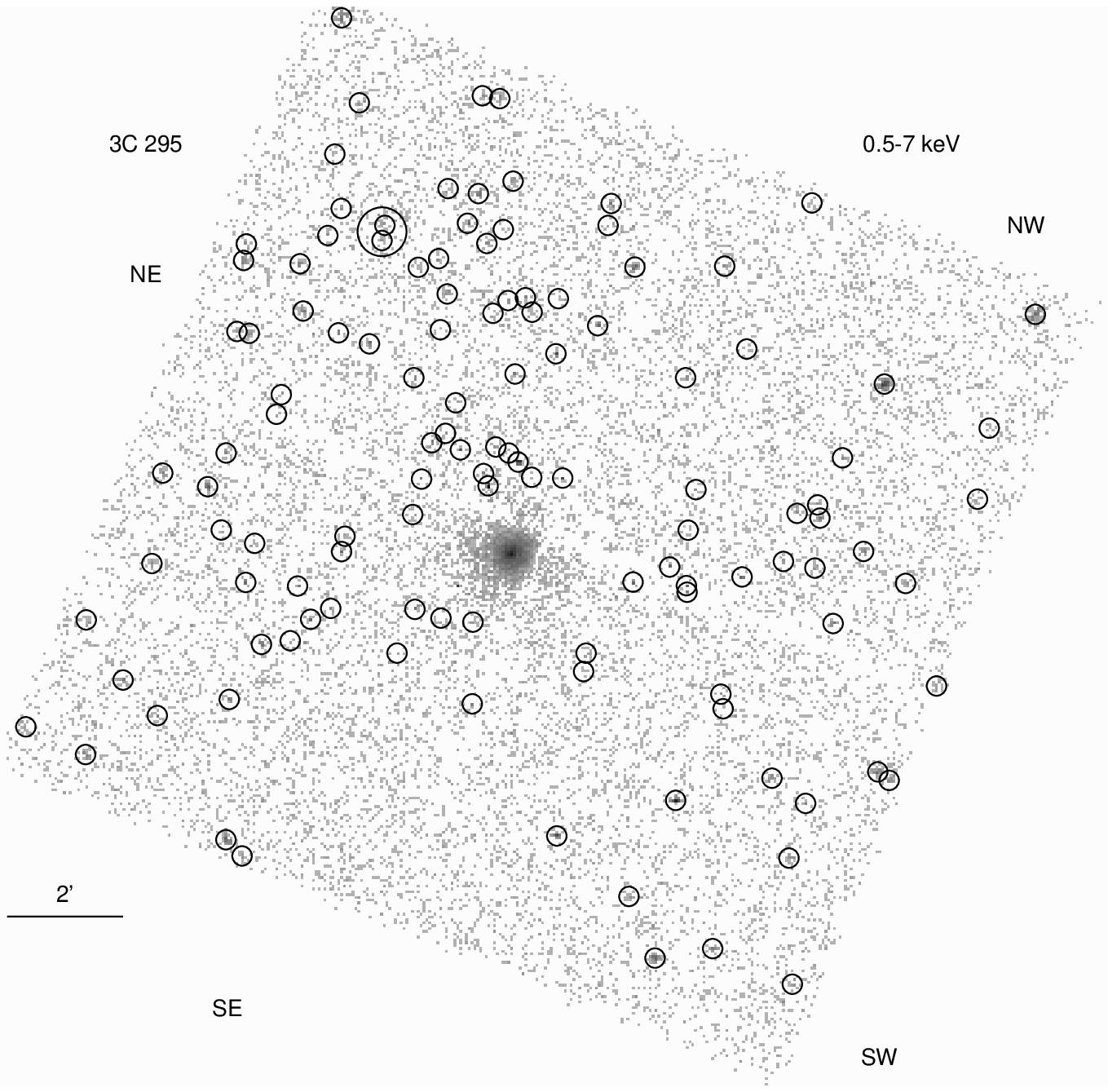}
\includegraphics[scale=0.30]{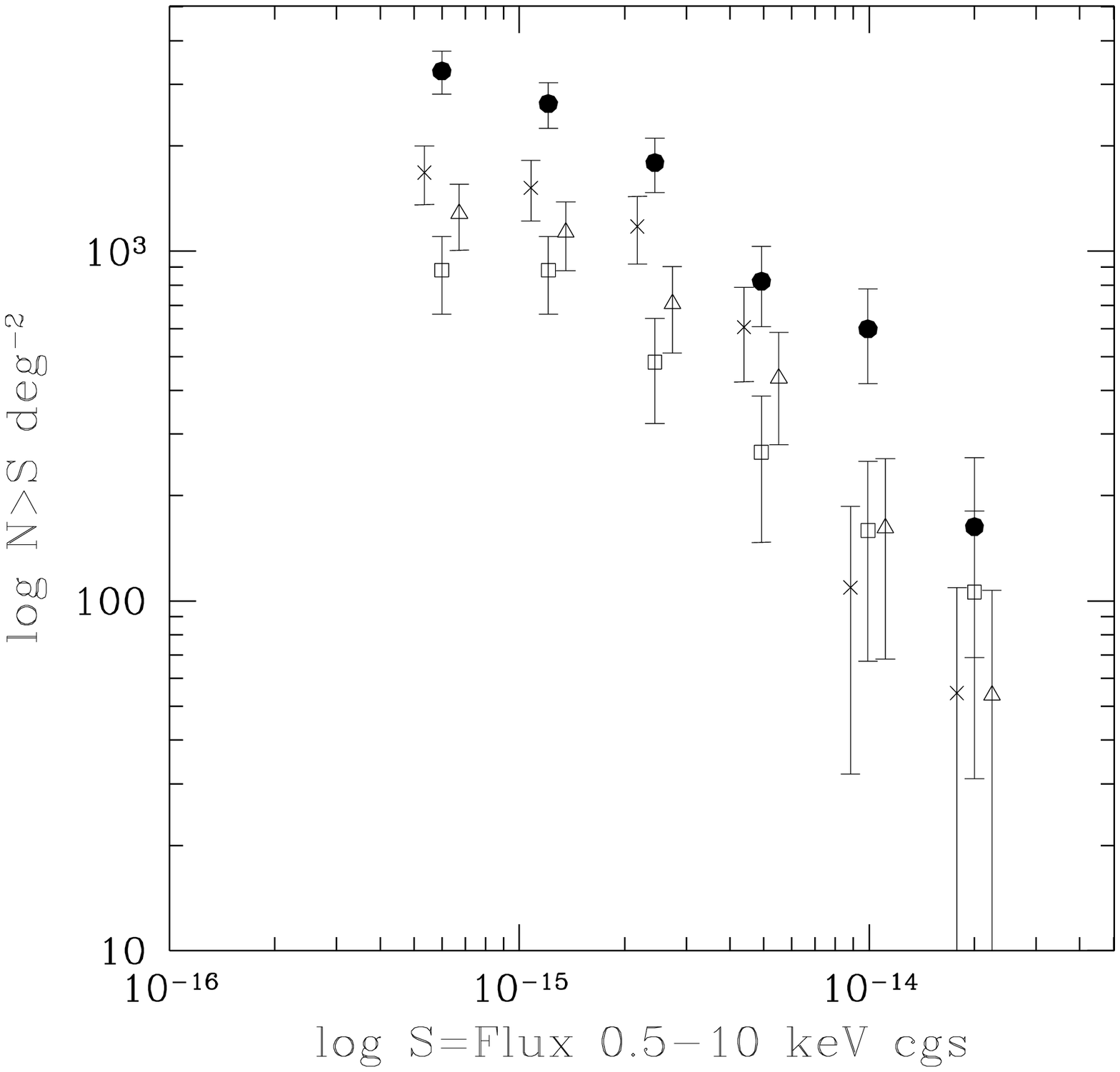}
\caption{Left: the Chandra 3C 295 field in the $0.5-7$ keV band. 
Circles represent the sources detected; the brightest source in the center of the field
is the cluster of galaxies 3C 295. Right: the mean (whole field) 3C 295 logN-logS in the
$0.5-10$ keV band, calculated for each ACIS-I chip
separately. Filled circles: NE chip, open triangles: NW, open squares: SE, crosses: SE.
Errors represent $1\sigma$ confidence limit.}
\end{figure}

\begin{figure}[ht]
\includegraphics[scale=0.30]{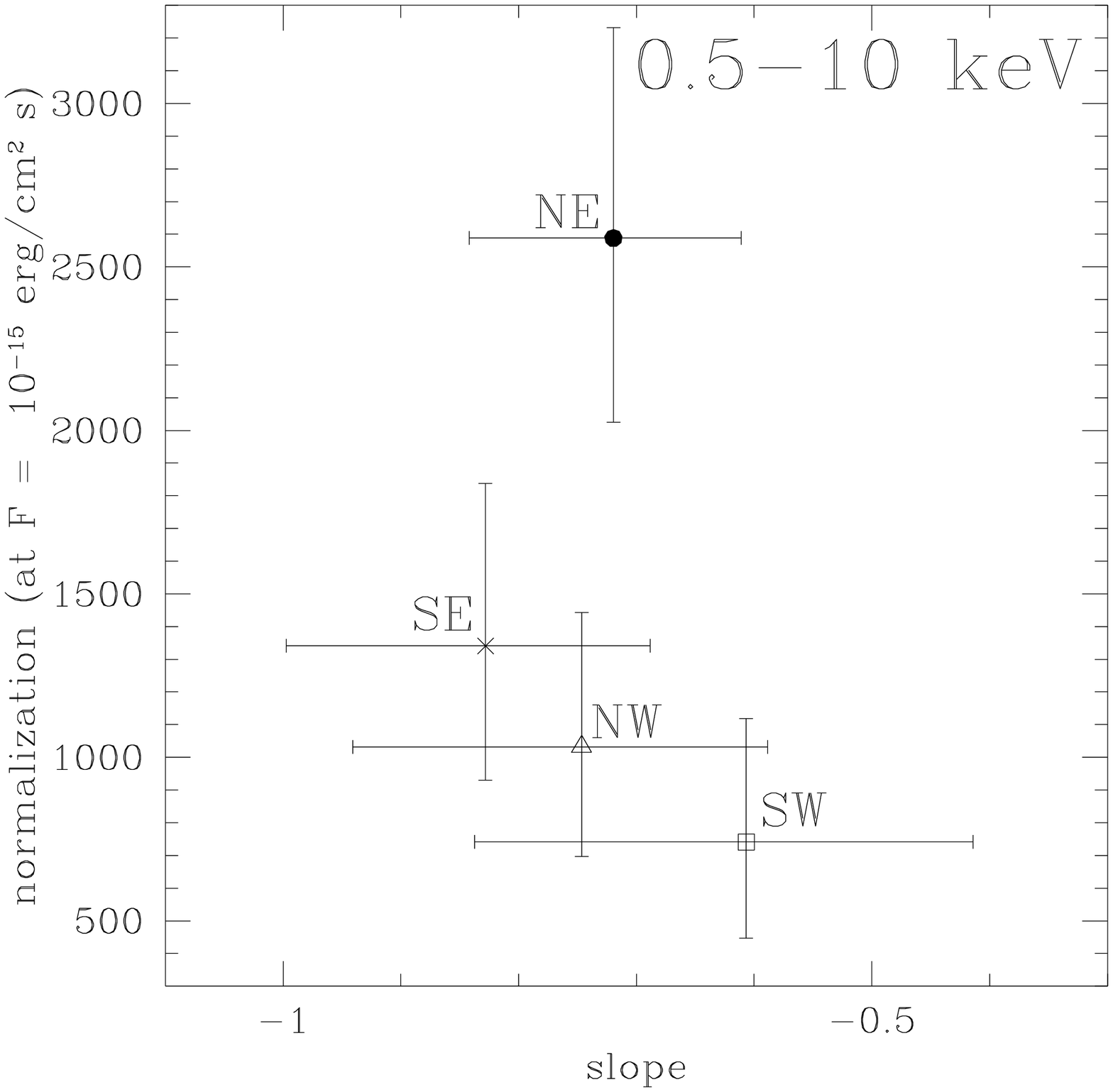}
\includegraphics[scale=0.30]{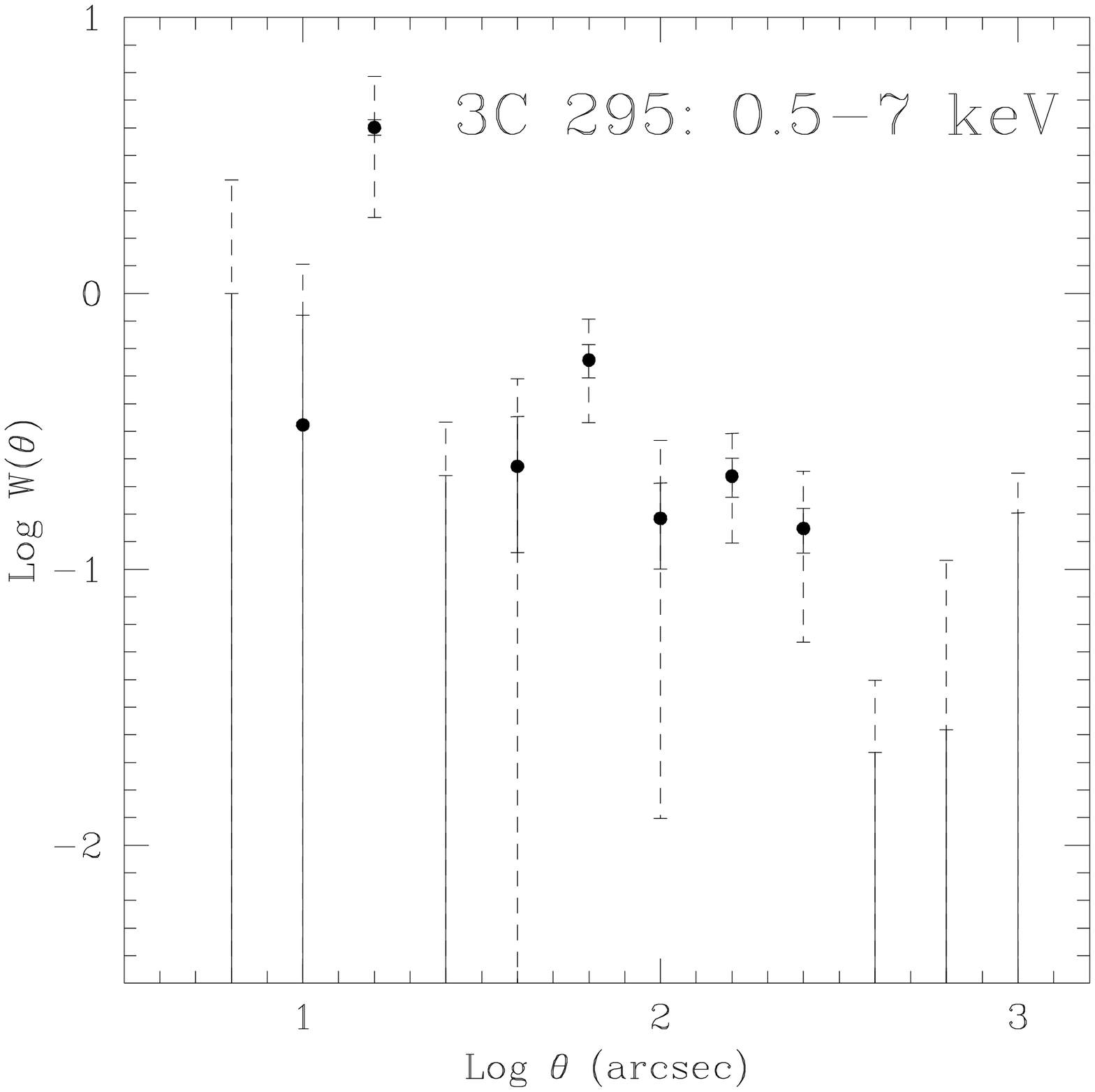}
\caption{ Left: results of the power law fits to the four logN-logS chips in
the $0.5-10$ keV band. Symbols refer to chips as in fig. 1b. Errors
are the $90\%$ confidence limit. 
Right: the 3C 295 ACF in the $0.5-7$ keV band. Solid error bars are Poisson;
dashed error bars are bootstrap.}
\end{figure}

\end{document}